\documentclass[aps,showpacs,amsmath,amssymb,twocolumn,nofootinbib]{revtex4-1}

\usepackage{subfigure}
\usepackage{graphicx}
\usepackage{dcolumn}
\usepackage{bm}
\usepackage{color}
\usepackage[colorlinks=true,pdfstartview=FitV,linkcolor=blue,citecolor=blue,urlcolor=blue]{hyperref}
\usepackage[mathlines]{lineno}
\usepackage[dvipsnames]{xcolor}
\usepackage{amsmath}

\everymath{\displaystyle}
\begin{document}

\newcommand{\up}[1]{$^{#1}$}
\newcommand{\down}[1]{$_{#1}$}
\newcommand{\powero}[1]{\mbox{10$^{#1}$}}
\newcommand{\powert}[2]{\mbox{#2$\times$10$^{#1}$}}

\newcommand{\evm}{\mbox{\rm{eV\,$c^{-2}$}}}
\newcommand{\mevm}{\mbox{\rm{MeV\,$c^{-2}$}}}
\newcommand{\gevm}{\mbox{\rm{GeV\,$c^{-2}$}}}
\newcommand{\pgd}{\mbox{g$^{-1}$\,d$^{-1}$}}
\newcommand{\um}{\mbox{$\mu$m}}
\newcommand{\spix}{\mbox{$\sigma_{\rm pix}$}}
\newcommand{\pav}{\mbox{$\langle p \rangle$}}

\newcommand{\sige}{\mbox{$\bar{\sigma_e}$}}
\newcommand{\mass}{\mbox{$m_\chi$}}
\newcommand{\crystal}{\mbox{$f_\textnormal{c}(q,E_e)$}}
\newcommand{\electron}{\mbox{$\rm{e^-}$}}

\newcommand{\beq}{\begin{equation}}
\newcommand{\eeq}{\end{equation}}
\newcommand{\beqs}{\begin{eqnarray}}
\newcommand{\eeqs}{\end{eqnarray}}

\title{Quintessence Axion Dark Energy and a Solution to the Hubble Tension}

\author{Gongjun Choi,$^{1}$}
\thanks{{\color{blue}gongjun.choi@gmail.com}}

\author{Motoo Suzuki,$^{1}$}
\thanks{{\color{blue}m0t@icrr.u-tokyo.ac.jp}}

\author{Tsutomu T. Yanagida,$^{1,2}$}
\thanks{{\color{blue}tsutomu.tyanagida@ipmu.jp}}

\affiliation{$^{1}$ Tsung-Dao Lee 
Institute, Shanghai Jiao Tong University, Shanghai 200240, China}
\affiliation{$^{2}$ Kavli IPMU (WPI), UTIAS, The University of Tokyo,
5-1-5 Kashiwanoha, Kashiwa, Chiba 277-8583, Japan}
\date{\today}

\begin{abstract}
We present a model in which the question about a nature of the dark energy and the recently raised Hubble tension can be addressed at once. We consider the electroweak axion in the minimal supersymmetric standard model where the axion energy density is identified with the observed dark energy. Along with this, imposing a gauged $Z_{10}$ symmetry makes it possible to have a gravitino dark matter whose mass amounts to $\sim\mathcal{O}(1)\,\,{\rm GeV}$. We find that the gravitino with mass $\sim\mathcal{O}(1)\,\,{\rm GeV}$ can be a good candidate of a decaying dark matter of which decay after recombination can reconcile discrepancy in local measurements of the Hubble expansion rate $H_{0}$ and that inferred from the cosmic microwave background observation.
\end{abstract}

\maketitle
\section{Introduction}  
One of the surprising observations in physics in the last century is the nonvanishing  dark energy (or cosmological constant) \cite{Perlmutter:1997zf,Schmidt:1998ys,Riess:1998cb}. If we take seriously a landscape conjecture based on the string theories \cite{Vafa:2005ui,Ooguri:2006in,Brennan:2017rbf,Dvali:2014gua,Dvali:2017eba,Sethi:2017phn,Danielsson:2018ztv,Obied:2018sgi,Ooguri:2018wrx}, the dark energy (DE) could be an almost static potential energy of a scalar field whose mass is about $10^{-33}$ eV \cite{Fukugita:1994xn,Frieman:1995pm,Choi:1999xn}. However, perturbative quantum corrections by gravitational interactions generate a huge mass for the scalar field even if it does not couple to any particle in the standard model sector. An interesting candidate for such a light scalar field will be a (pseudo) Nambu-Goldstone (NG) boson, since the perturbative gravitational corrections never violate global symmetry at least. However, the absence of any exact global symmetry is suggested by quantum gravity~\cite{Hawking:1987mz,Lavrelashvili:1987jg,Giddings:1988cx,Coleman:1988tj,Gilbert:1989nq,Banks:2010zn} and hence the relevant global symmetry must have explicit breaking terms generating a mass of the NG boson. One simple example for such an explicit breaking is a coupling to an anomaly term of some non-Abelian gauge field. This thought motivated Fukugita and one of us (T.T.Y.) to introduce a pseudo NG boson generating the DE \cite{Fukugita:1994xn}.\footnote{In fact, almost the present value of the DE $\sim$ $(3\pm1 {\rm meV})^4$ was introduced in \cite{Fukugita:1994xn} to compensate an inconsistency between the stellar age and the measured Hubble constant $H_{0}$ even before the observation of the DE \cite{Perlmutter:1997zf,Schmidt:1998ys,Riess:1998cb}.} We call this pseudo NG boson as a quintessence axion.
Interesting is that we can explain the observed dark energy $\sim$ (1m eV)$^4$ if the quintessence axion couples to the known electroweak (EW) SU(2) gauge instanton \cite{Nomura:2000yk,Ibe:2018ffn}. 

On the other hand, disagreement between local measurements of the expansion rate of the universe ($H_{0}$) \cite{Riess:2016jrr,Riess:2018byc,Bonvin:2016crt,Birrer:2018vtm} and that from the cosmic microwave background (CMB) observation \cite{Aghanim:2018eyx} has exceeded $\sim4\sigma$ level and it may be signaling a feature of a new physics. Among many different suggestions for resolving the tension, a decaying dark matter (DDM) solution implies a dark matter with a life time longer than the age of the universe \cite{Vattis:2019efj}. The decay of DDM may start after recombination and DM abundance today and the amount of DE in the model may differ from those from CMB data assuming $\Lambda$CDM cosmology.

In this letter, given the mysterious two puzzles in the standard cosmology described thus far, we present a model in which we may be able to address both question simultaneously. We consider an EW $SU(2)_{L}$ axion within the minimal supersymmetric standard model (MSSM). We argue that a gravitino and a slowly rolling EW axion field serve as a DDM and a quintessence field for DE, respectively. Moreover, the quanta of the EW axion and its superpartner, axino, take the role of product particles resulting from decay of the gravitino DDM. We find that the model is able to achieve the correct amount of DE and DDM mass in a consistent manner provided that we assume an anomaly-free $Z_{10}$ flavor symmetry.

\section{Model}
\label{sec:model}
Based on the MSSM, we introduce one pseudo Nambu-Goldstone chiral superfield  $\mathcal{A}$ whose imaginary part of the complex boson component is the axion ($a$). The theory is assumed to have an invariance under the shift of $\mathcal{A}$, that is, $\mathcal{A}\rightarrow\mathcal{A}+i\alpha$  except for the EW $SU(2)_{L}$ gauge anomaly term ($\alpha$ is a real constant). In the MSSM, there is an accidental global symmetry $U(1)_{B+L}$ besides the shift symmetry which is, however, anomalous for $SU(2)_{L}$. Thus, we introduce higher dimensional operators to break the $U(1)_{B+L}$ explicitly so that we can generate a mass for the EW $SU(2)_{L}$ axion \cite{Nomura:2000yk}.

The axion superfield coupling at low energy is given by,
\beq
\mathcal{L}_{\rm eff}=\int d^2\theta\frac{1}{32\pi^2}\frac{\mathcal{A}}{F_\mathcal{A}}\mathcal{W}\mathcal{W} + {\rm h.c.},
\eeq
where $\mathcal{W}$ is $SU(2)_L$ gauge field strength and $F_\mathcal{A}$ is the decay constant of the axion. Hereafter, we omit the gauge and spinor indices for simplicity. We take $F_\mathcal{A}\sim M_{P} \sim 2.4 \times 10^{18}$\,GeV so that the quintessence mechanism naturally works \cite{Nomura:2000yk}, where $M_{P}$ is the reduced Planck mass.

\begin{figure}[t]
\centering
\hspace*{-5mm}
\includegraphics[width=0.47\textwidth]{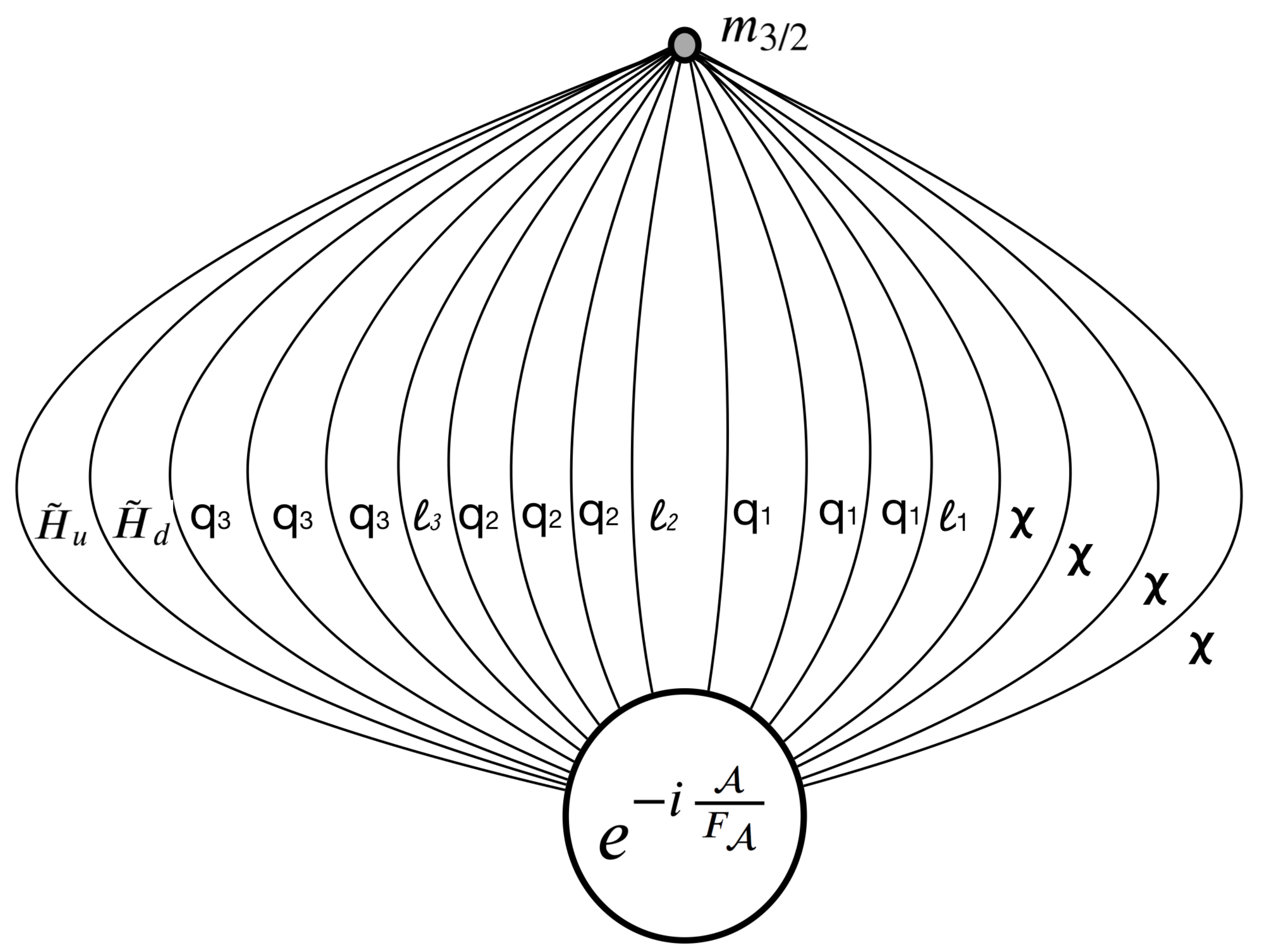}
\caption{One anti-instanton diagram generating the axion potential $e^{-i\mathcal{A}/F_\mathcal{A}}$. Together with the higher dimension operator $m_{3/2}\prod_{i=1}^{3}(q_iq_iq_il_i)(\chi\chi)^2(\tilde H_u\tilde H_d)$, we obtain the axion potential given in Eq.~(\ref{eq:DEscale}).}
\vspace*{-1.5mm}
\label{fig:1}
\end{figure}

The potential of the axion is generated by the EW $SU(2)_{L}$ instantons and the dynamical scale of the potential is calculated%
\footnote{We impose a $U(1)_R$ symmetry as in Table 1 of Ref.~\cite{Nomura:2000yk}. } as~\cite{Choi:1996fs,Choi:1998ep,Nomura:2000yk}
\beq
\label{eq:lambda}
\Lambda^4\sim c\, e^{-\frac{2\pi}{\alpha_2(M_{P})}}m_{3/2}^3 M_{P},
\eeq
where $\alpha_2(M_{P})$ is the $SU(2)_L$ gauge coupling constant at the Planck scale, $m_{3/2}$ denotes the gravitino mass, and $c$ is the model dependent constant which is discussed in the following. In order to suppress dangerous dimension 5 operators for the proton decay $\tilde{\mathcal{O}}=QQQL$ in the superpotential \cite{Sakai:1981pk,Weinberg:1981wj}, an Abelian flavor symmetry $U(1)_{F}$ was introduced under which the quarks and leptons are charged \cite{Nomura:2000yk}. In this case, the numerical constant $c$ becomes extremely small due to the suppression by high powers of $U(1)_{F}$ breaking parameter, i.e., $c\simeq 10^{-13}$ \cite{Nomura:2000yk}.%
\footnote{The $U(1)_F$ breaking parameter $\epsilon$ can be determined to be $\epsilon\simeq1/17$ to explain the quark and lepton mass matrices~\cite{Yanagida:1998jk,Ramond:1998hs}. Then, in the instanton calculus, the coefficient $c$ in Eq.~(\ref{eq:lambda}) is estimated as $\epsilon^{10}\simeq 10^{-13}$ to close all the fermion zero modes.}
Provided the EW axion plays the role of quintessence field for the DE, the gravitino mass $m_{3/2}\simeq\mathcal{O}(1){\rm TeV}$ is required to explain the observed DE, i.e., $\Lambda^{4}_{\rm DE}\sim(1{\rm meV})^{4}$ by the axion potential in Eq.~(\ref{eq:lambda}).

On the other hand, if we impose a discrete gauge symmetry as the flavor symmetry rather than the continuous $U(1)_{F}$, we obtain a drastically different gravitino mass. As a matter of fact, the discrete $Z_{10}$ is anomaly free \cite{Ibe:2018ffn} with the charge assignment done in Table 2 of Ref.~\cite{Nomura:2000yk}. Thus, we assume the anomaly free gauged $Z_{10}$ symmetry to suppress the dangerous dimension 5 operators for the proton decay. Then, all fermion zero modes are closed by inserting one higher dimensional operator (see Fig.~\ref{fig:1}),\footnote{We thanks to M. Ibe and M. Yamazaki for discussion about this in the private communication.}
\beq
\mathcal{L}=\kappa\, m_{3/2}^2m_{3/2}^\dagger\prod_{i=1}^{3}(q_iq_iq_il_i)(\chi\chi)^2(\tilde H_u\tilde H_d),
\eeq
where $q_i$ and $l_i$ denote the quarks and leptons of three families, $\chi$ is the $SU(2)_L$ gaugino, $\tilde H_{u,d}$ are the higgsinos, $\kappa$ is an unknown constant which we assume $\kappa\simeq \mathcal{O}(1)$. Here and Hereafter, we take a unit of $M_{P}=1$ unless otherwise specified. Notice that one insertion of $m_{3/2}$ is necessary to make the operator consistent with $U(1)_{R}$ symmetry. The coefficient of $m_{3/2}m_{3/2}^\dagger$ comes from the superspace integration of the K{\"a}hler potential~\cite{Choi:1996fs,Choi:1998ep}. The total flavor charge of this operator is zero. Eventually, we obtain
\beq
\Lambda^4\simeq\left(\frac{\kappa}{10^{-4}}\right) \left(\frac{m_{3/2}}{1\,{\rm GeV}}\right)^3(1\times 10^{-3}\,{\rm eV})^4\,,
\label{eq:DEscale}
\eeq
where $\alpha_{2}(M_{P})=1/23$ was used.\footnote{The weak-gravity conjecture \cite{ArkaniHamed:2006dz} is satisfied for $F_{\mathcal{A}}=M_{P}$ if $\alpha(M_{P})\simeq2\pi$ holds. This condition is easily achieved by introducing massive matter particles at intermediate energy scales. It is surprising that the condition in Eq.~(\ref{eq:DEscale}) does not change due to a miraculous SUSY cancellation as shown in \cite{Ibe:2018ffn}.} Now it becomes clear that matching the EW axion energy scale in Eq.~(\ref{eq:DEscale}) to the cosmological constant requires $m_{3/2}\sim(0.1-1)$\,GeV for $\kappa=1-10^{-4}$. With the decay constant $F_{\mathcal{A}}\sim M_{P}$, we obtain the axion mass $m_a\lesssim 10^{-33}$\,eV which is less than the current Hubble expansion rate. Indeed, such an EW axion is able to serve as a slowly rolling quintessence dark energy field.  

\begin{figure}[t]
\centering
\hspace*{-5mm}
\includegraphics[width=0.47\textwidth]{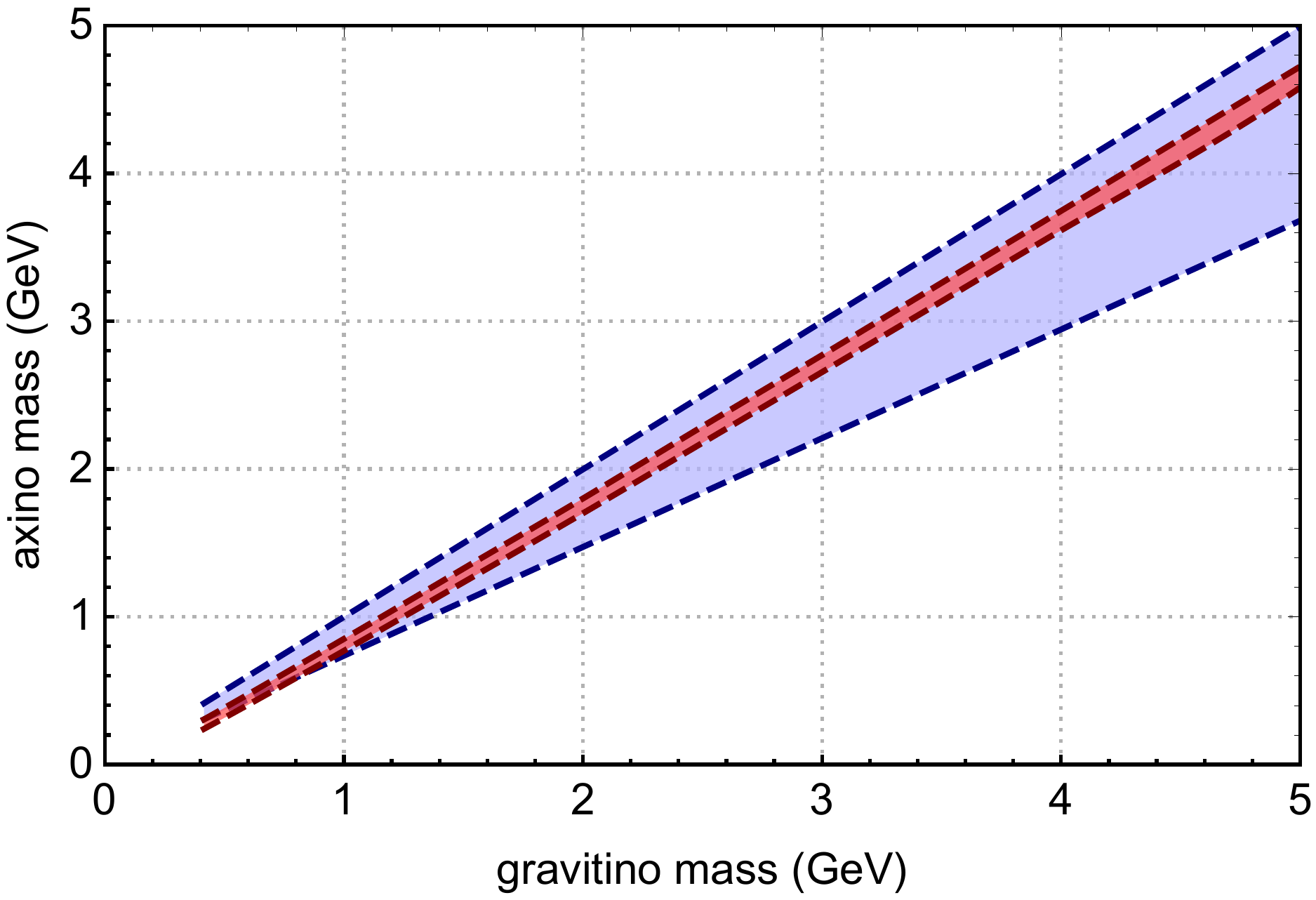}
\caption{Allowed mass range for the gravitino parent DM and axino daughter massive particle near $m_{3/2}\simeq\mathcal{O}(1){\rm GeV}$, which is obtained by mapping constraints on $\epsilon$ and $\tau$ at 68\% C.L. presented in \cite{Vattis:2019efj}. The blue region is obtained from the constraint on $\epsilon$ parameter while the red region is from the constraint on $\tau$. The overlapping region is understood as the eventual allowed region for resolving the Hubble tension.}
\vspace*{-1.5mm}
\label{fig:2}
\end{figure}

\section{Reconciling Hubble Tension}
\label{sec:hubble}
In this section, we examine how the model presented in the previous section can help us to reconcile the Hubble tension. Our model is a concrete particle physics model for the decaying dark matter resolution to the Hubble tension suggested in \cite{Vattis:2019efj} and therefore its parameter space should be subject to constraints in \cite{Vattis:2019efj}.\footnote{For another example of particle physics model for decaying dark matter resolution to the Hubble tension, see, e.g. Refs.~\cite{Choi:2020tqp}.}

The basic strategy taken in \cite{Vattis:2019efj} for resolving the Hubble tension is to make evolution of the Hubble expansion rate $H(z)$ after recombination different from that in $\Lambda$CDM cosmology so that $H_{0}$ obtained in $\Lambda$DDM becomes greater than that from $\Lambda$CDM. To this end, a Monte Carlo Markov Chain (MCMC) was performed based on $\Lambda$DDM model with four free parameters with priors and several data points for the values of Hubble expansion rate at different redshifts within $0\leq z\leq2.4$ were used. The four free parameters include a fraction of a parent DM rest mass transferred to a daughter massless particle ($\epsilon$), a life time of the parent DM ($\tau$), a dark matter abundance today $\Omega_{{\rm DM}}$ and a reduced Hubble parameter $h=H_{0}$/(100km/sec/Mpc) which are used to infer the parent DM energy density at recombination via $\rho_{{\rm DDM}}(a_{{\rm rec}})=\Omega_{{\rm DM}}\rho_{{\rm crit}}a^{-3}_{{\rm rec}}$. In the $\Lambda$DDM model, $\rho_{{\rm DDM}}(a_{{\rm rec}})a_{{\rm rec}}^{3}$ value starts to decrease after onset of decay of DDM instead of remaining conserved in time. This results in an earlier transition from matter dominated era to DE dominated era. 

The DDM decay produces a massless and a massive daughter particles of which four momenta are given as $p_{\mu}=(\epsilon m_{{\rm DDM}}, \overrightarrow{p})$ and $p_{\mu}'=((1-\epsilon) m_{{\rm DDM}},- \overrightarrow{p})$, respectively. Interestingly, it was shown in \cite{Vattis:2019efj} that the massive daughter particle is still distinguished from an ordinary matter in that its equation of state deviates from zero. With the framework described above, $\Lambda$DDM model succeeded in showing that the modified evolution of the Hubble expansion rate can ease the Hubble tension for the reported parameter spaces of the four free parameters aforementioned.

Given the constraints on the free parameters in \cite{Vattis:2019efj}, we can study how those can be applied to the physical picture we presented in Sec.~\ref{sec:model}. For our model, we consider a scenario where the gravitino takes the role of DDM of which decay results in two products including an EW axion and its fermionic superpartner, axino. The former is regarded as a massless particle which inherits the energy of $\epsilon\,m_{3/2}$ from the gravitino while the later serves as a massive warm daughter particle. Now we go through mapping of the constraints on the four free parameters in \cite{Vattis:2019efj} to constraints on the gravitino and axino mass, and the reheating temperature below.

Firstly, we notice that the constraint on $-2.88\leq\log_{10}\epsilon\leq-0.64$ (68\% C.L.) in \cite{Vattis:2019efj} can be converted into the constraint on $m_{\tilde{a}}/m_{3/2}$ via the dispersion relation of the axino ($\tilde{a}$)
\beq
E_{\tilde{a}}^{2}=m_{\tilde{a}}^{2}+|\overrightarrow{p}|^{2}\,\,\,\leftrightarrow\,\,\,(1-\epsilon)^{2}m_{3/2}^{2}=m_{\tilde{a}}^{2}+\epsilon^{2}m_{3/2}^{2}\,.
\eeq
With this, we apply the constraint on the lifetime of DDM in \cite{Vattis:2019efj}, i.e.,  $1.3\leq\log_{10}(\tau/{\rm Gyr})\leq2.18$ (68\% C.L.), to the following decay rate of gravitino ($\tilde{\Psi}_{\mu}$) \cite{Hamaguchi:2017ihw}
\beq
\Gamma(\tilde{\Psi}_{\mu}\rightarrow\tilde{a}+a)=\frac{m^{3}_{3/2}}{192\pi M_{P}^{2}}(1-r_{m})^{2}(1-r_{m}^{2})^{3}\,.
\label{eq:gravitino_gamma}
\eeq
Then, we obtain a constraint on $m_{3/2}$. In Eq.~(\ref{eq:gravitino_gamma}), $m_{3/2}$ and $m_{\tilde{a}}$ are the gravitino and axino mass respectively and $r_{m}\equiv m_{\tilde{a}}/m_{3/2}$ is used. In Fig.~\ref{fig:2}, we show the allowed parameter space for the gravitino and axino mass so obtained for $m_{3/2}$ near $\mathcal{O}(1){\rm GeV}$. The blue and red region is based on the constraints on $\epsilon$ and $\tau$, respectively. The overlapping region is understood as the eventual allowed region for $(m_{3/2},m_{\tilde{a}})$ to resolve the Hubble tension. The full allowed gravitino mass to resolve the Hubble tension ranges from $\mathcal{O}(0.1){\rm GeV}$ to $\mathcal{O}(1){\rm TeV}$. Intriguingly, we observe that the $m_{3/2}$ range in Fig.~\ref{fig:2} covers the gravitino mass range capable of reproducing the scale of the dark energy via Eq.~(\ref{eq:DEscale}).\footnote{Another different DDM model was proposed based on the QCD axion in \cite{Hamaguchi:2017ihw} by referring to \cite{Berezhiani:2015yta,Chudaykin:2016yfk,Poulin:2016nat}. There, DM population consists of three components including the gravitino, the axino and the axion. With a life time shorter than the age of universe, the minor component decay to the major component and the radiation. As the minor component, the gravitino NLSP around $m_{3/2}\sim\mathcal{O}(1){\rm 
GeV}$ was discussed, but without a compelling reason for the mass besides a resolution to the Hubble tension. In our scenario, there is only one single component of DM population, which decay to a massive and massless particle with a lifetime greater than the age of universe.} 

\begin{figure}[t]
\centering
\hspace*{-5mm}
\includegraphics[width=0.47\textwidth]{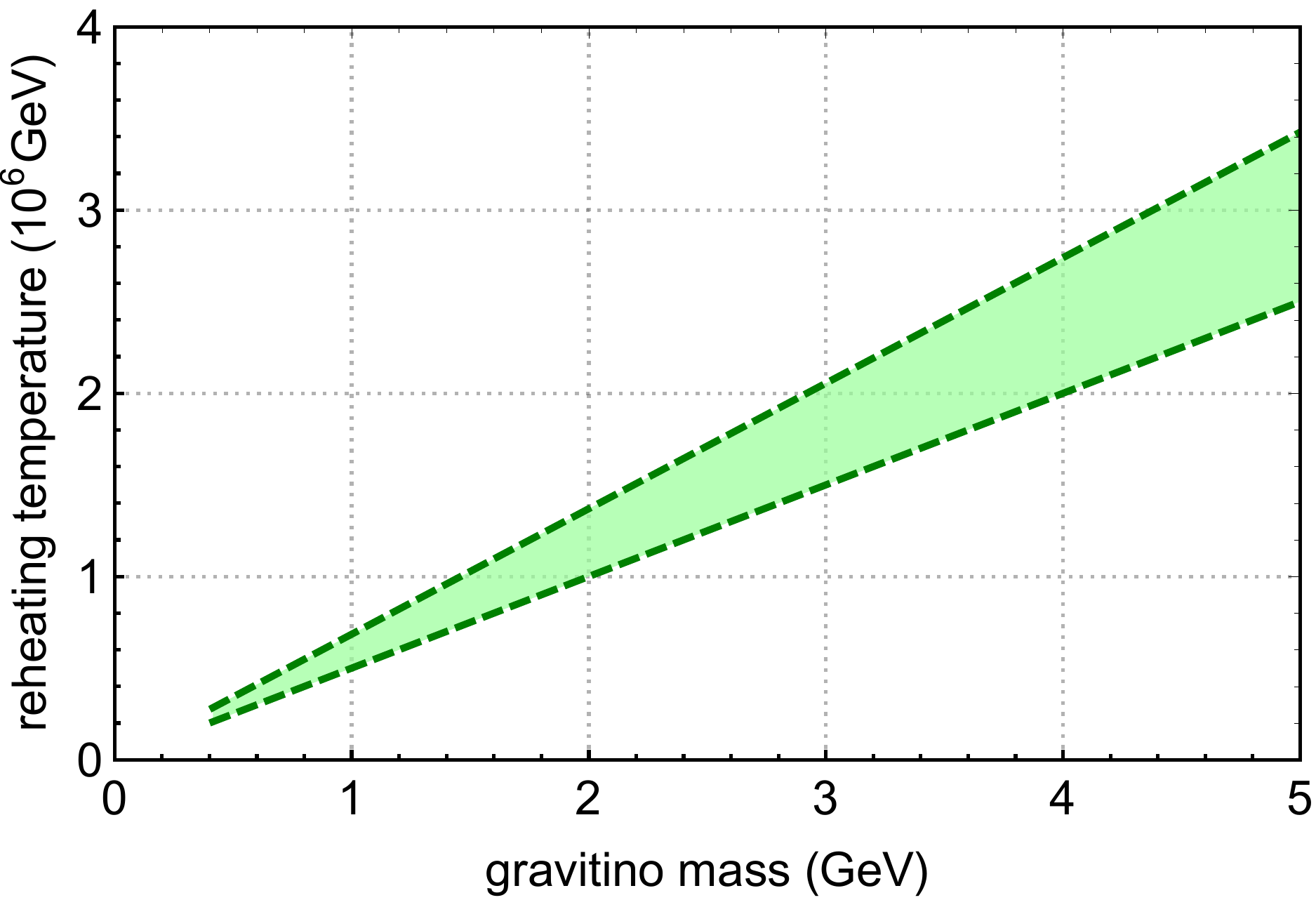}
\caption{Allowed parameter space for the reheating temperature obtained from constraints on $\Omega_{{\rm DM}}$ and $h$ in \cite{Vattis:2019efj}. The plot is shown for the gravitino mass range of our interest near $\sim1{\rm GeV}$.}
\vspace*{-1.5mm}
\label{fig:3}
\end{figure}

Secondly, the individual constraints on $\Omega_{{\rm DM}}$ and $h$ in \cite{Vattis:2019efj} gives the constraint on $\Omega_{{\rm DM}}h^{2}$, which is $0.099\leq\Omega_{{\rm DM}}h^{2}\leq0.137$ (68\% C.L.). Application of this constraint to the following gravitino DM abundance today \cite{Hamaguchi:2017ihw,Bolz:2000fu,Pradler:2006qh,Rychkov:2007uq}
\beqs
\Omega_{3/2}h^{2}&\simeq&
0.2\times\left(\frac{T_{R}}{10^{6}{\rm GeV}}\right)\times\left(\frac{1{\rm GeV}}{m_{3/2}}\right)\cr\cr
&&\times\left(\frac{M_{3}(T_{R})}{3{\rm TeV}}\right)^{2}\times\left(\frac{\gamma(T_{R})/(T^{6}_{R}/M_{P}^{2})}{0.4}\right)\,,
\label{eq:gravitino_abundance}
\eeqs
yields a constraint on the reheating temperature for a range of the gravitino mass. In Eq.~(\ref{eq:gravitino_abundance}), $T_{R}$ is the reheating temperature, $M_{3}$ is the running gluino mass and $\gamma$ is the gravitino production rate. With exemplary values of $M_{3}\simeq3{\rm TeV}$ and $\gamma(T_{R})/(T^{6}_{R}/M_{P}^{2})\simeq0.4$, we show in Fig.~\ref{fig:3} the allowed parameter space for the reheating temperature so obtained for the gravitino DM mass range of our interest near $\sim1{\rm GeV}$. Remarkably, we realize that the required reheating temperature to accomplish the thermal production of the gravitino mass near $\sim1{\rm GeV}$ can be consistent with the non-thermal leptogenesis \cite{Fukugita:1986hr,Buchmuller:2005eh}.

Within the picture we discussed so far, one may wonder whether the saxion (the real part of the complex boson component of the chiral superfield $\mathcal{A}$) can form the other component of DDM than the gravitino. In order to guarantee that the gravitino is the only DDM candidate in the model, we should suppress the primordial production of the relic saxion from its coherent oscillation.\footnote{If the initial amplitude of the saxion field is $\sim\mathcal{O}(M_{P})$, we have too large saxion density.} For that, we impose a discrete $Z_{2}$ symmetry under which $\mathcal{A}$ is odd \cite{Takahashi:2005kp}\footnote{The K{\"a}hler potential of the axion superfield is a function of $(\mathcal{A}+\mathcal{A}^{\dagger})^{2n}$ ($n=1,2,3,..$).} and assume that the induced mass of saxion from its coupling to inflaton is larger than the Hubble expansion rate during inflation \cite{Enqvist:1987au}. On top of this, it is expected that the thermal production of saxion and axino is highly suppressed as well due to the decay constant $F_{\mathcal{A}}$ comparable to $M_{P}$. Thereby, the model contains the gravitino with $m_{3/2}\sim\mathcal{O}(1){\rm GeV}$ as the sole candidate of DDM.  

\section{Conclusions}
In this letter, we have pointed out possible candidates of DE and DM within MSSM with a chiral superfield $\mathcal{A}$ for the EW axion. The model contains $U(1)_{R}\times Z_{10}$ symmetry and the shift symmetry of the $\mathcal{A}$ field. The $Z_{10}$ flavor symmetry is necessary to suppress higher dimensional operators $\tilde{\mathcal{O}}=QQQL$ dangerous for the proton decay. 

Now encountering the problems for the nature of DE and the recently raised Hubble tension, we addressed the problems in this letter by (1) imposing a gauged $Z_{10}$ flavor symmetry and (2) taking the decay constant of the EW axion to be $F_{\mathcal{A}}\simeq M_{P}$. These enable us to obtain (1) the dynamical scale of the EW axion potential comparable to the current DE density $\sim(1{\rm meV})^{4}$, (2) the EW axion mass around $\sim10^{-33}\!-\!10^{-34}{\rm eV}$ and (3) the gravitino mass $\sim\mathcal{O}(1){\rm GeV}$. Therefore, we could identify the EW axion as a quintessence field for the DE. Also, by converting the constraints on the $\Lambda$DDM model parameters in \cite{Vattis:2019efj} to those on the gravitino mass, axino mass and reheating temperature, we showed that the gravitino with $m_{3/2}\sim\mathcal{O}(1){\rm GeV}$ can be a candidate of DDM with the EW axion and axino as the decay products. With such a small mass of the gravitino, the most natural SUSY breaking mediation mechanism to the MSSM sector is the gauge mediation \cite{Dine:1981za,Dimopoulos:1981au,Dine:1981gu,Dine:1982zb,Nappi:1982hm,AlvarezGaume:1981wy,Dimopoulos:1982gm,Dine:1993yw,Dine:1995ag,Dine:1994vc}. Finally, in this letter we have constructed a $\Lambda$DDM model reconciling the Hubble tension assuming the quintessence axion model. However, it is easily extended to a QCD axion model with a larger decay constant like a string axion model with $F_{\mathcal{A}}\simeq 10^{16}{\rm GeV}$.

Finally, let us comment on the small-scale problems in the cold dark matter paradigm. In Ref.~\cite{Bae:2019vyh},%
\footnote{See also Refs.~\cite{Peter:2010au,Peter:2010jy,Bell:2010fk,Peter:2010sz,Aoyama:2011ba,Wang:2012eka,Wang:2013rha,Cheng:2015dga}.} 
a similar setup to ours is discussed as a solution to the small-scale problems (especially too-big-to-fail problem), where the axino dark matter with the lifetime $\simeq 10$\,Gyr decays into the slightly lighter gravitino with the kick velocity $(m_{\tilde a}-m_{3/2})/m_{ 3/2}\simeq 10^{-4}c$ and the axion. Contrary to this, we find that the decaying gravitino DM discussed in our work is characterized by the longer life time ($\sim35{\rm Gyr}$) and the larger kick velocity $\sim10^{-3}c-10^{-1}c$ and thus irrelevant to the small scale problems.


\begin{acknowledgments}
G. C. thanks to Robert Shrock for drawing attention to the Hubble tension problem. T. T. Y. thanks to Danny Marfatia for introducing the reference \cite{Vattis:2019efj}. T. T. Y. is supported in part by the China Grant for Talent Scientific Start-Up Project and the JSPS Grant-in-Aid for Scientific Research No. 16H02176, No. 17H02878, and No. 19H05810 and by World Premier International Research Center Initiative (WPI Initiative), MEXT, Japan. T. T. Y. thanks to Hamamatsu Photonics. 

\end{acknowledgments}


\bibliography{ref}

\end{document}